\documentclass[a4paper,11pt]{article}
\usepackage{pos}
\usepackage[group-separator={,}]{siunitx} 
\usepackage{enumitem} 
\usepackage{physics}

\title{Performance optimizations for porting the openQ$^\star$D package to GPUs}

\author*[a]{Roman Gruber}
\author[b]{Anton Kozhevnikov}
\author[a]{Marina Krstić Marinković}
\author[a,b]{Thomas C. Schulthess}
\author[b]{Raffaele Solcà}

\affiliation[a]{D-PHYS, ETH Zürich,\\ Zürich, Switzerland}
\affiliation[b]{Swiss National Supercomputing Centre,\\ Lugano, Switzerland}

\emailAdd{rgruber@ethz.ch}
\emailAdd{anton.kozhevnikov@cscs.ch}
\emailAdd{marinama@phys.ethz.ch}
\emailAdd{schulthess@cscs.ch}
\emailAdd{raffaele.solca@cscs.ch}

\def\code#1{\texttt{#1}}

\def\df#1{\textbf{\textit{#1}}}

\usepackage{floatrow}
\DeclareFloatFont{small}{\scriptsize}
\definecolor{codegreen}{RGB}{0, 153, 0}
\definecolor{codegray}{RGB}{127, 127, 127}
\definecolor{codeblue}{RGB}{102, 214, 237}
\definecolor{codekeyword}{RGB}{249, 36, 114}
\definecolor{codecomment}{RGB}{127, 127, 127}
\definecolor{backcolor}{RGB}{242, 242, 235}
\definecolor{linkcolor}{RGB}{102, 0, 0}
\definecolor{corange}{RGB}{255, 70, 0}
\definecolor{cyellow}{RGB}{209, 153, 0}
\definecolor{cblue}{RGB}{64, 128, 255}
\definecolor{cbrown}{RGB}{153, 102, 51}
\definecolor{cpink}{RGB}{255, 0, 255}
\definecolor{cred}{RGB}{255, 64, 0}
\definecolor{cgreen}{RGB}{0, 191, 0}
\definecolor{clightblue}{RGB}{191, 217, 255}
\definecolor{cturquois}{RGB}{0, 255, 255}
\definecolor{cpurple}{RGB}{128, 0, 255}
\definecolor{clightgreen}{RGB}{175, 255, 175}
\definecolor{clightpink}{RGB}{255, 175, 255}
\definecolor{cdarkblue}{RGB}{0, 0, 255}
\definecolor{cdarkred}{RGB}{255, 0, 0}
\definecolor{cdarkgreen}{RGB}{0, 255, 0}

\abstract{OpenQ$^\star$D code has been used by the RC$^\star$ collaboration for the generation of fully dynamical QCD+QED gauge configurations with C$^\star$ boundary conditions. In this talk, optimization of solvers provided with the openQ$^\star$D package relevant for porting the code on GPU-accelerated supercomputing platforms is discussed. We present the analysis of the current implementations of the GCR solver preconditioned with Schwarz alternating procedure for ill-conditioned Dirac-operators. With the goal of enabling support for GPUs from various vendors, a novel method of adaptive CPU/GPU-hybrid implementation is proposed.}

\FullConference{%
 The 38th International Symposium on Lattice Field Theory, LATTICE2021
  26th-30th July, 2021
  Zoom/Gather@Massachusetts Institute of Technology
}

\begin{document}
\maketitle

\section{Introduction}

\label{sec:intro}

The advent of GPUs in modern supercomputers enables the path towards exascale computing, where the peak operations per second is around $10^{18}$ \cite{top500}. To reach such a peak performance is challenging and highly depends on the problem as well as the involved data types. Since lattice QFT calculations are bound by memory bandwidth\footnote{One of the most important kernels is the application of the Dirac operator to a spinor field, which can be seen as a variant of sparse matrix-vector multiplication (SpMV).} and not by compute performance, one must think about how to reduce memory traffic in order to increase performance. For example, data types with smaller bit lengths can be considered.

The scope of this work is an analysis of the different solver algorithms used in the lattice QFT application \code{openQ*D-1.1} \cite{openqxd}. This software package is used for the generation of fully dynamical QCD+QED gauge configurations with C$^\star$ boundary conditions and $O(a)$-improved Wilson-fermions. Different aspects of the solvers are highlighted to find potential for improvement.

The analysis in this document is performed using Python-implementations of the examined kernels. This switch of programming language and philosophy enabled to run the kernels with simulated data types that are usually non-accessible within the native application without significant implementation effort.

\section{Conjugate Gradient}

\label{sec:cg}

The conjugate gradient kernel \code{cgne()}\footnote{See line 429ff in \code{modules/linsolv/cgne.c} in \cite{openqxd}} implements the algorithm already in mixed precision. The complete  kernel was simulated using different data types -- floats as well as posits\footnote{To produce the plots, the Dirac operator \code{Dop\_dble()} was extracted in binary64 format from the original code running a simulation of a $4^4$ lattice, Schrödinger Functional (SF) boundary conditions (\code{type 1}), no C$^\star$ boundary conditions (\code{cstar 0}) and $1$ rank. The first 2000 trajectories were considered thermalisation. The matrix was extracted in trajectory 2001. A Python script mimicking the exact behaviour of the \code{cgne()} kernel from the source code, was implemented to cope with arbitrary data types.}. The simulated data types were binary64, binary32, tensorfloat32, binary16, bfloat16, posit32, posit16, and posit8 (please refer to table \ref{tab:limits} for more information on these formats). The considered Dirac-operator represented as a CSR-matrix had approximately 2\% non-zero values.

\begin{figure}[htbp]
    \centering
    \includegraphics[width=1.0\textwidth, clip=true, trim = 0 325 0 0]{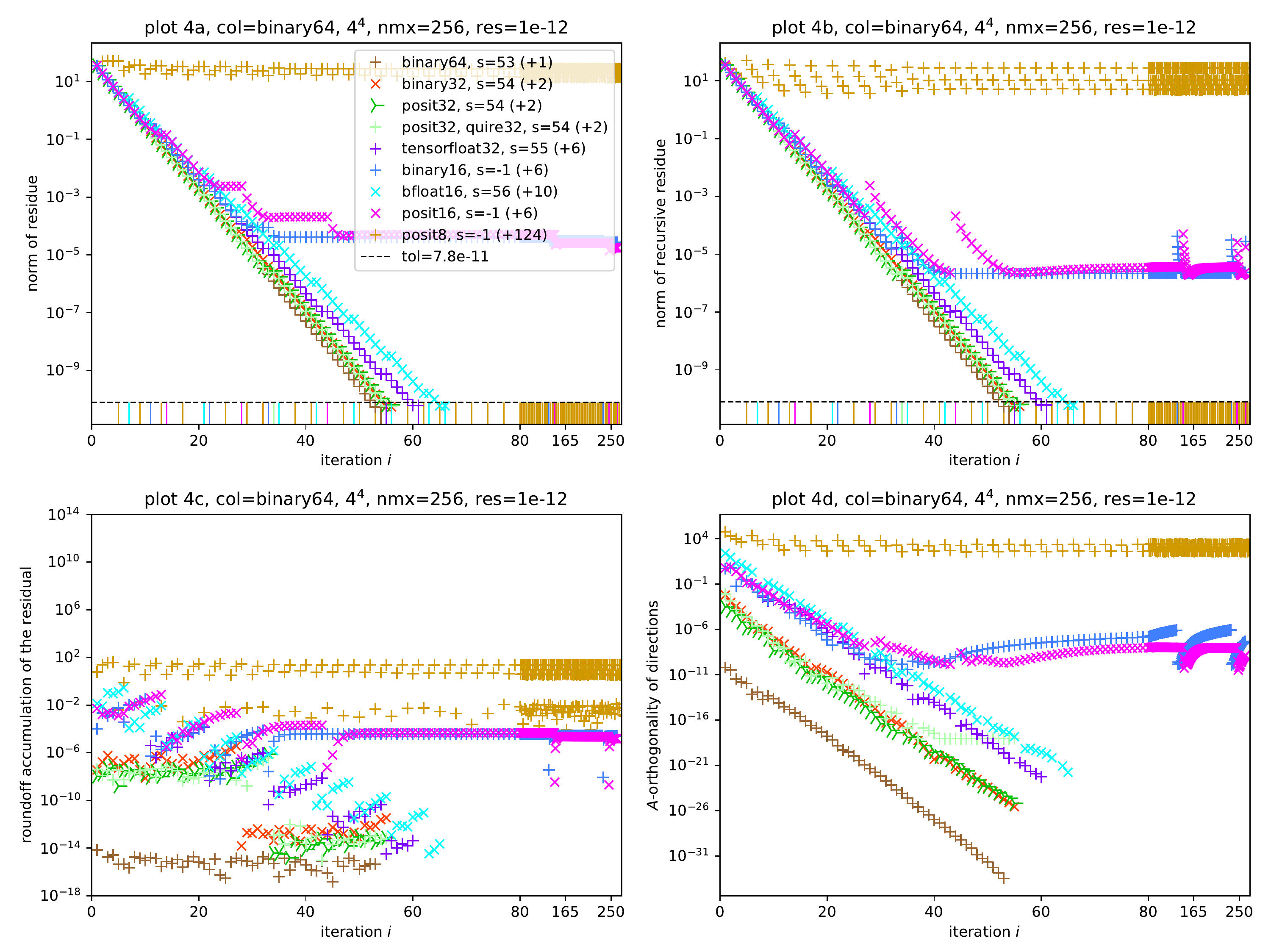}
    \caption{Convergence analysis of a conjugate gradient run, where binary32 was replaced by one of the simulated data types. The number \code{s} describes the number of regular CG-steps needed (the value of \code{status}), while the number in the brackets indicate the number of reset steps. Plot \textit{4a} shows the exact residue, $\vec{r}_i = \vec{b} - A\vec{x}_i$, calculated in every iteration in binary64, while plot \textit{4b} shows the norm of the recursively calculated residue, $\vec{r}_i = \vec{r}_{i-1} + \alpha_{i-1} A \vec{p}_{i-1}$, (cast to binary64 after calculation). The algorithm executes reset steps when the residue is lowered by an amount roughly the machine epsilon of the datatype (cf. table \ref{tab:limits}). These are indicated at the bottom of the plot. The relative residue suffers from round-off accumulation because of its recursive calculation; this is the difference between the two residues in plots \textit{4a} and \textit{4b}, which is plotted in plot \textit{4c}. Plot \textit{4d} shows the $A$-orthogonality of the current direction with respect to the last direction, namely the value of $\vec{p}_{i-1}^\dagger A \vec{p}_{i}$. The value of \code{res} -- the desired relative residue of the calculated solution -- is set to $10^{-12}$.}
    \label{fig:cgne}
\end{figure}

Figure \ref{fig:cgne} shows all the simulated data types using a reduction data type of binary64, meaning that all reduction operations where conducted in binary64\footnote{Eg: norms}. The following hierarchy is expected (smaller means convergence in fewer steps):

\begin{align}
    \textcolor{cbrown}{\text{binary64}} < \textcolor{cgreen}{\text{posit32}} \leq \textcolor{cred}{\text{binary32}} \leq \textcolor{cpurple}{\text{tensorfloat32}} \leq \textcolor{cturquois}{\text{(1)}} \leq \textcolor{cpink}{\text{posit16}} \leq \textcolor{cblue}{\text{binary16}} \leq \textcolor{cturquois}{\text{(2)}} < \textcolor{cyellow}{\text{posit8}}, \label{eq:hierarchy}
\end{align}

where \textcolor{cturquois}{\text{bfloat16}} could be either at position \textcolor{cturquois}{\text{(1)}} or \textcolor{cturquois}{\text{(2)}}, depending on what is more important; precision or number range.

Notice that hte target relative residual, $10^{-12}$, is outside the representable number range of binary16, posit16 and posit8. These data types cannot reach the target tolerance, therefore we didn't expect them to converge. This is indeed the case. Furthermore, we see that binary16 and posit16 both are not able to go below $10^{-5}$, leading to no further considerable progress after step 45. This can be seen by the recursive residue stalling or even increasing -- an indicator that the data type has reached its limits.

Both, binary32 and posit32, required the same number of steps, although round-off accumulation and $A$-orthogonality are slightly better for posit32. The reason for this is due to the higher density of posits in the relevant number regime (between $-1$ and $1$) leading to higher precision.

Finally, we compare the three data types with the same exponent range, but different precisions; binary32, tensorfloat32 and bfloat16 (cf. table \ref{tab:limits}). The less precision, the slower the convergence. The price to go from 23 to 10 mantissa bits results in 1 more conjugate gradient step as well as 4 more reset steps. When going further down to 7 mantissa bits again 1 more regular step and 4 more reset steps where needed to finally bring bfloat16 to convergence after 56 regular plus 10 reset steps. Bearing in mind that it occupies only 16 bits, this is a remarkable result, way better than its 16-bit competitors.

As seen in plot \textit{4a}, all data types start to converge by the same speed (all slopes are equal). The least precise data type, namely bfloat16 with its 7 mantissa bits, resets first, followed by binary16 and tensorfloat32, both with 10 mantissa bits. The next one is posit16, because it has more precision than binary16 in the relevant regime, followed by binary32 with 23 mantissa bits and later by posit32, where the same argument as before holds. The curve of binary64 would also reset at some point, but that never triggered in this run.

Specially plot \textit{4a} suggests that we can start to calculate in a data type with 16 bits of length until we fall below a constant value (proportional to the machine epsilon), then continuing the calculation in a data type with 32 bit-length until that number regime is exhausted as well, again switching to a 64 bit data type to finish the calculation.

\subsection{Preliminary conclusions for the CG solver}

Reduction variables should be chosen in a data type with large precision and number range, such as  binary64, regardless of the current data type, since type conversions between different IEEE floating-point types are not considered to be expensive. On the other hand, the number of variables needed in that data type does not scale with the problem size or the number of steps, we can use a data type with large bit-length such as binary64.

The difference between tensorfloat32, binary32 and bfloat16 answers the question how important precision is in the calculation. The only relevant difference was in the number of reset steps. If the data type is lower in bit-length, the memory-boundedness of the problem suggests that the calculation performs faster. The trade-off is the amount of (computationally more expensive) reset steps that increase with lower precision.

\bgroup
\def\arraystretch{1.2}
\begin{table}
\centering

    \begin{tabular}{ |p{3cm}||p{1.8cm}|p{1.8cm}|p{1.8cm}|p{1.8cm}|p{1.8cm}|  }
        \hline
        \multicolumn{6}{|c|}{Floating-point format limits} \\
        \hline
        data type & $f_{max}$ & $f_{min}$ & $f_{smin}$ & sign. digits \footnotemark & machine $\epsilon$ \\
        \hline
        binary64 (e=11, m=52)
            & \num[round-mode = figures, round-precision = 2, scientific-notation = true]{1.79769313486231570e308}
            & \num[round-mode = figures, round-precision = 2, scientific-notation = true]{2.22507385850720138e-308}
            & \num[round-mode = figures, round-precision = 2, scientific-notation = true]{4.94065645841246544e-324}
            & $\le 15.9$
            & \num[round-mode = figures, round-precision = 2, scientific-notation = true]{2.2204460492503131e-16} \\
        binary32 (e=8, m=23)
            & \num[round-mode = figures, round-precision = 2, scientific-notation = true]{3.402823466e38}
            & \num[round-mode = figures, round-precision = 2, scientific-notation = true]{1.175494350e-38}
            & \num[round-mode = figures, round-precision = 2, scientific-notation = true]{1.401298464e-45}
            & $\le 7.2$
            & \num[round-mode = figures, round-precision = 2, scientific-notation = true]{1.1920928955078125e-07} \\
        binary16 (e=5, m=10)
            & \num[round-mode = figures, round-precision = 2, scientific-notation = true]{65504}
            & \num[round-mode = figures, round-precision = 2, scientific-notation = true]{0.00006103515625}
            & \num[round-mode = figures, round-precision = 2, scientific-notation = true]{5.9604644625e-8}
            & $\le 3.3$ 
            & \num[round-mode = figures, round-precision = 2, scientific-notation = true]{0.0009765625} \\
        bfloat16 (e=8, m=7)
            & \num[round-mode = figures, round-precision = 2, scientific-notation = true]{3.38953138e38}
            & \num[round-mode = figures, round-precision = 2, scientific-notation = true]{1.175494350e-38}
            & \num[round-mode = figures, round-precision = 2, scientific-notation = true]{9.18354961e-41}
            & $\le 2.4$ 
            & \num[round-mode = figures, round-precision = 2, scientific-notation = true]{0.0078125} \\
        tfloat32 (e=8, m=10)
            & \num[round-mode = figures, round-precision = 2, scientific-notation = true]{3.40116213e38}
            & \num[round-mode = figures, round-precision = 2, scientific-notation = true]{1.175494350e-38}
            & \num[round-mode = figures, round-precision = 2, scientific-notation = true]{1.14794370e-41}
            & $\le 7.2$
            & \num[round-mode = figures, round-precision = 2, scientific-notation = true]{0.0009765625} \\
        \hline
        posit32 (es=2)
            & \num[round-mode = figures, round-precision = 2, scientific-notation = true]{1.32922799578491587290e36}
            & \num[round-mode = figures, round-precision = 2, scientific-notation = true]{7.52316384526264005099e-37}
            & N/A
            & $\le 8.1$
            & \num[round-mode = figures, round-precision = 2, scientific-notation = true]{7.450580596923828e-09} \\
        posit16 (es=1)
            & \num[round-mode = figures, round-precision = 2, scientific-notation = true]{2.68435456e8}
            & \num[round-mode = figures, round-precision = 2, scientific-notation = true]{3.72529029e-9}
            & N/A
            & $\le 3.6$
            & \num[round-mode = figures, round-precision = 2, scientific-notation = true]{0.000244140625} \\
        posit8 (es=0)
            & \num{64}
            & \num[round-mode = figures, round-precision = 2, scientific-notation = true]{0.015625}
            & N/A
            & $\le 1.5$
            & \num[round-mode = figures, round-precision = 2, scientific-notation = true]{0.03125} \\
        \hline
    \end{tabular}
    
    \caption{\label{tab:limits} Summary of highest representable numbers, $f_{max}$, minimal subnormal, $f_{smin}$, and non-subnormal, $f_{min}$, representable numbers above 0 in any common IEEE 754 floating-point and posit format \cite{ieee754_1985, bfloat16, tf32, posit2018standard} together with their approximated precision in decimal. $e$ and $m$ denote the number of exponent and mantissa bits for floats, whereas $es$ denotes the maximal number of exponent bits for posits.}
    
\end{table}
\egroup

\footnotetext{Number of significant digits in decimal.}

\section{Schwarz Alternating Procedure}

\label{sec:sap}

Domain decomposition is a way to partition the large system into (possibly many) smaller sub-problems with regularly updated boundary conditions coming from solutions of neighbouring sub-problems. They fit well into the notion of parallel processing because the sub-problem can be chosen to be contained in one single rank. The full lattice is split into sub-lattices called \df{local lattice}. Each rank has its own local lattice, the size of which is determined at compilation time. The full lattice consists of the ensemble of all local lattices arranged in a grid. These local lattices can be split as well into \df{blocks}. It is therefore advisable to choose the size of the blocks as divisor of the local lattice size such that one or more blocks fit into one rank. These sub-problems can then be solved using an iterative solving method.

\begin{figure}[htbp]
  \centering
  \includegraphics[width=0.3\textwidth]{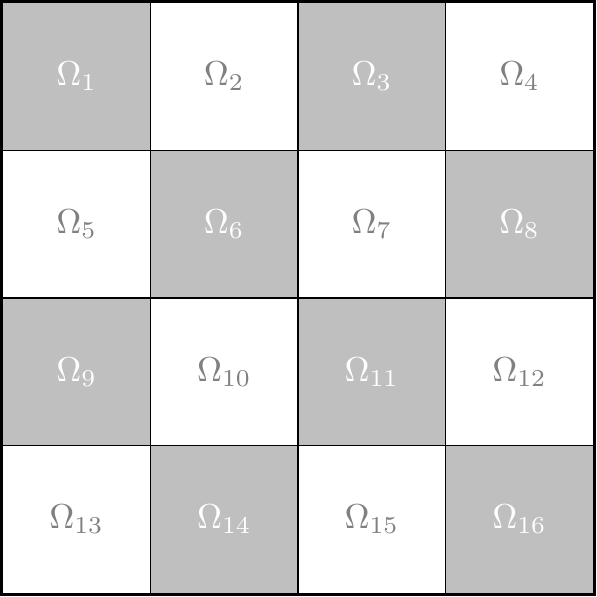}
  \caption{A $d=2$ dimensional example of a decomposition of a lattice $\Omega = \bigcup_{i=1}^{n} \Omega_i$ into $n=16$ domains named $\Omega_i$. Notice such a decomposition can always be colored like a chessboard.}
  \label{fig:ddecomp}
\end{figure}

The idea behind Schwarz Alternating Procedure (SAP) is to loop through all blocks $\Omega_i$ and solve the smaller sub-problem using boundary conditions given from the most recent global solution (cf. figure \ref{fig:ddecomp}). If the original problem only includes nearest-neighbour interactions, the solution of a block $\Omega_i$ depends only on that block and its exterior boundary points, which are the adjacent points on the neighbouring blocks with opposite color. For example, the solution of the sub-problem involving $\Omega_6$, depends only on the solutions of $\Omega_2$, $\Omega_5$, $\Omega_7$ and $\Omega_{10}$\footnotemark.
\footnotetext{It depends on all other sub-problems as well, but indirectly.}
Therefore, all grey (white) blocks can be solved simultaneously, with the most recent boundary conditions obtained from the white (grey) blocks. Solving all grey, followed by all white blocks is called a \df{Schwarz-cycle} and is considered one iteration in SAP. Each block can be solved with any desired solver separately.

\begin{figure}[htbp]%
    \centering
    \includegraphics[width=0.45\textwidth, clip=true, trim = 0 0 445 0]{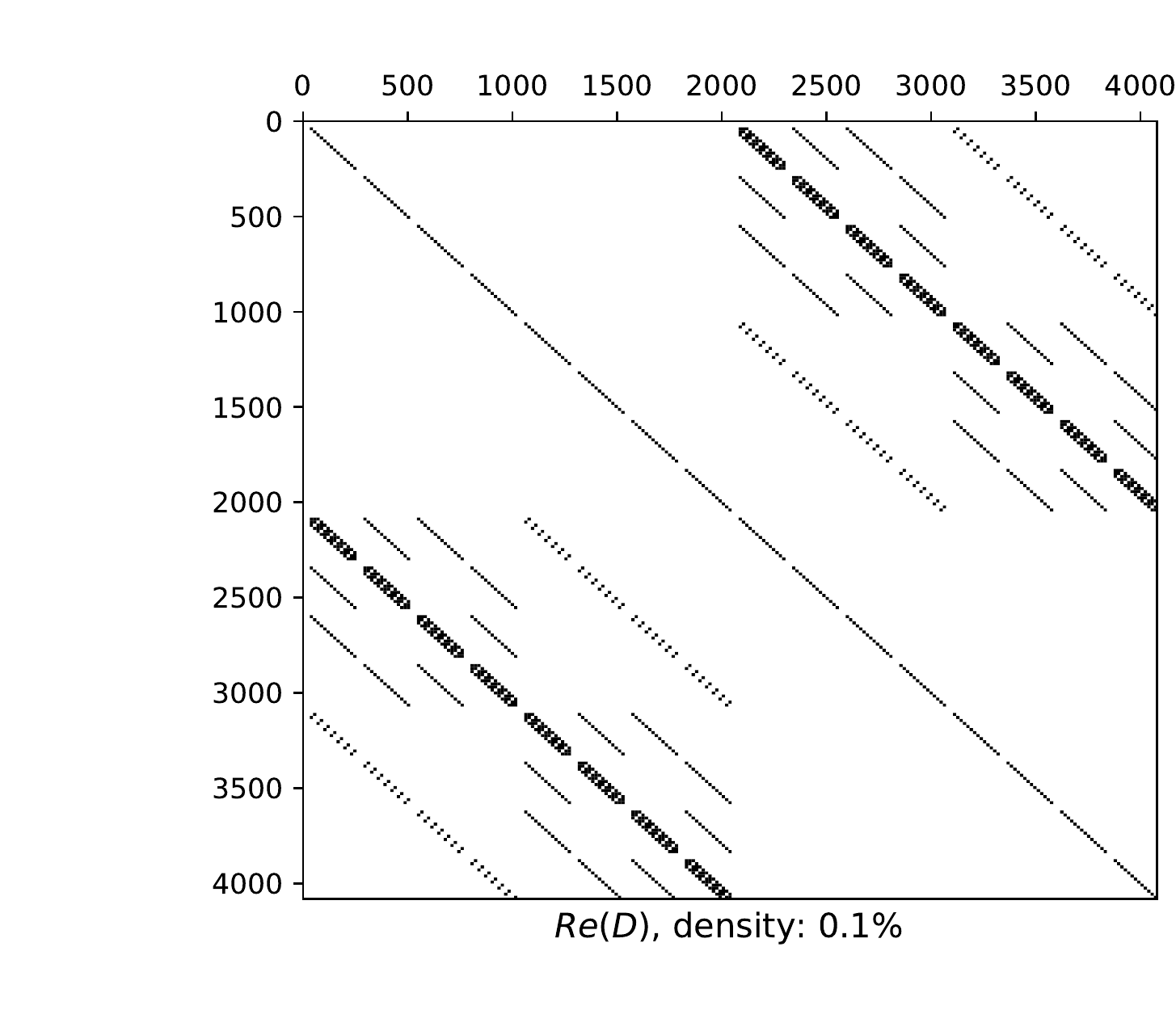} 
    \qquad
    \includegraphics[width=0.45\textwidth]{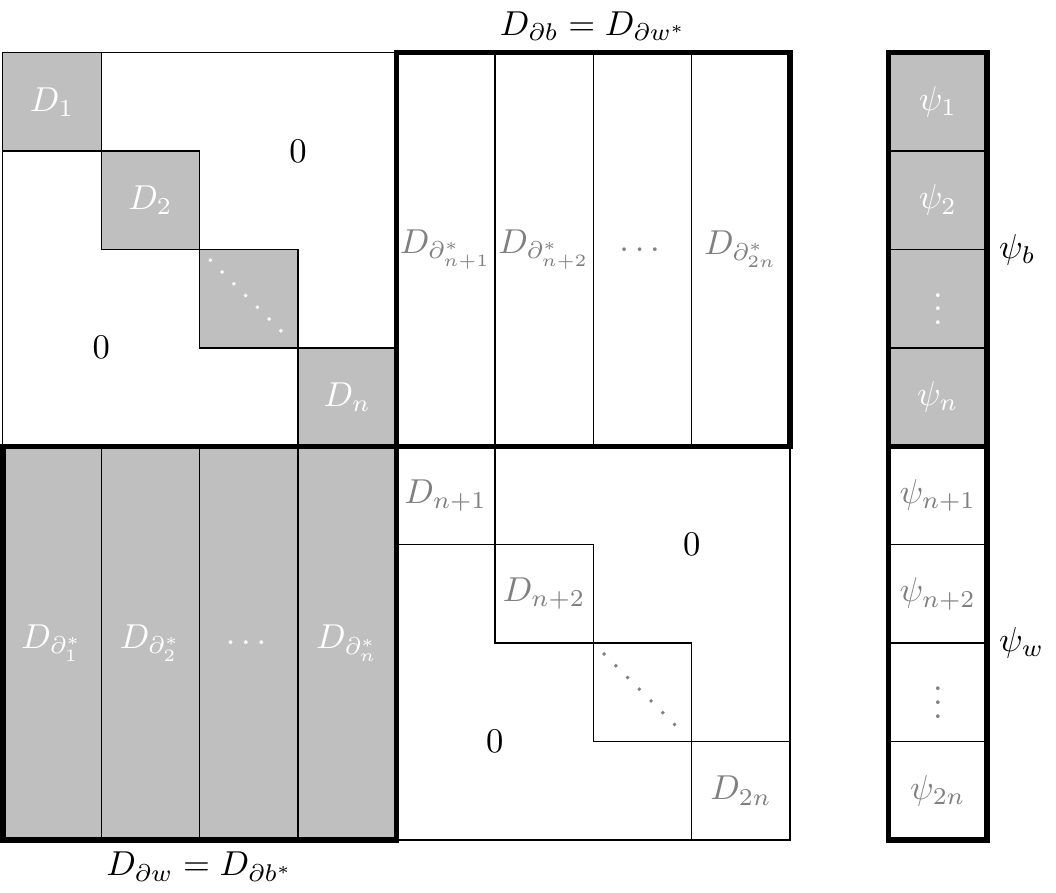} 
    \caption{\textit{Left}: An example plot of a Dirac-matrix of an $8^4$-lattice with SF-boundary conditions. The operator is already in a shape, where the even lattice points come first, followed by the odd lattice points. Every pixel consists of $192 \times 192$ real numbers. White pixels denote zero, black pixels non-zero. \textit{Right}: Schematic of the Dirac-operator in terms of a large sparse matrix. If the components of the black blocks are arranged such that they appear first, then the decomposition from figure \ref{fig:ddecomp} can be translated into a matrix with blocks as in the picture. $D_i$ describes the Dirac-operator restricted to block $i$ and $D_{\partial b}$ ($D_{\partial w}$) is the Dirac-operator restricted to the external boundaries of the black (white) blocks. The color external boundary operators can be decomposed into external boundary operators of the $i$-th block, $D_{\partial^{*}_i}$. The right side describes a vector decomposed into the same $2n$ domains $\psi_1, \dots, \psi_{2n}$. The upper half corresponds to the black blocks and the lower half to the white blocks.}%
    \label{fig:dirac_matrix}%
\end{figure}

Whereas the division into domains on the lattice is straightforward, the representation of the Dirac-operator as a sparse matrix and its decomposition is not. Looking at an actual example of a Dirac-operator written as a matrix (cf. figure \ref{fig:dirac_matrix} left), one observes a lot of structure: while on the diagonal we find the operators restricted to the black and white blocks, the first and the third quadrant describe the operators restricted to the interior and exterior boundaries. The decomposition into $2n$ domains ($n$ grey and $n$ white blocks) can be translated as seen in figure \ref{fig:dirac_matrix} right. Notice that the restricted operators $D_i$ are easily solved, because they have block diagonal form.

\subsection{Setup}

The complete SAP+GCR kernel was implemented using Python in the same way as the \code{fgcr()} function from the source code\footnote{See line 212ff in \code{modules/linsolv/fgcr.c} in \cite{openqxd}.}. The Python implementation allowed a floating-point data type for the reduction variables separately (\code{rdtype}). It also accepts a "large" data type (\code{ldtype}) by which the restart steps are calculated in, and a "small" data type (\code{sdtype}) in which the regular and the MR steps are performed in. The result is obtained in terms of the "large" data type. There are various configuration settings to choose from (cf. table \ref{tab:sap_gcr_settings}).

\begin{table}[htbp]
\centering
    \begin{tabular}{ |p{1.5cm}|p{10cm}|  }
        \hline
        setting & meaning \\
        \hline\hline
        \code{res}  & desired relative residual \\
        \hline
        \code{nmx}  & maximal number of GCR steps \\
        \hline
        \code{nkv}  & number of generated Krylov vectors until restarting the algorithm \\
        \hline
        \code{ncy}  & number of SAP-cycles to perform in each iteration \\
        \hline
        \code{nmr}  & number of MR-steps to perform on each block in each SAP-cycle \\
        \hline
        \code{bs}   & block size \\
        \hline
        \code{ldtype}  & "large" data type \\
        \cline{0-1}
        \code{rdtype}  & reduction data type \\
        \cline{0-1}
        \code{sdtype}  & "small" data type \\
        \hline
    \end{tabular}
    \caption{Settings for SAP+GCR and their meanings.}
    \label{tab:sap_gcr_settings}
\end{table}

The possible data types for \code{ldtype}, \code{rdtype} and \code{sdtype} are binary64 and binary32\footnote{Unfortunately, there was no possibility to use binary16, bfloat16 or tensorfloat32, even though modern GPUs such as the one tested on do support these data types. The reason for this is the data types were not available in the used CUDA library, CuPy \cite{cupy2017}.}.

\begin{figure}[htbp]
    \centering
    \includegraphics[width=1.0\textwidth]{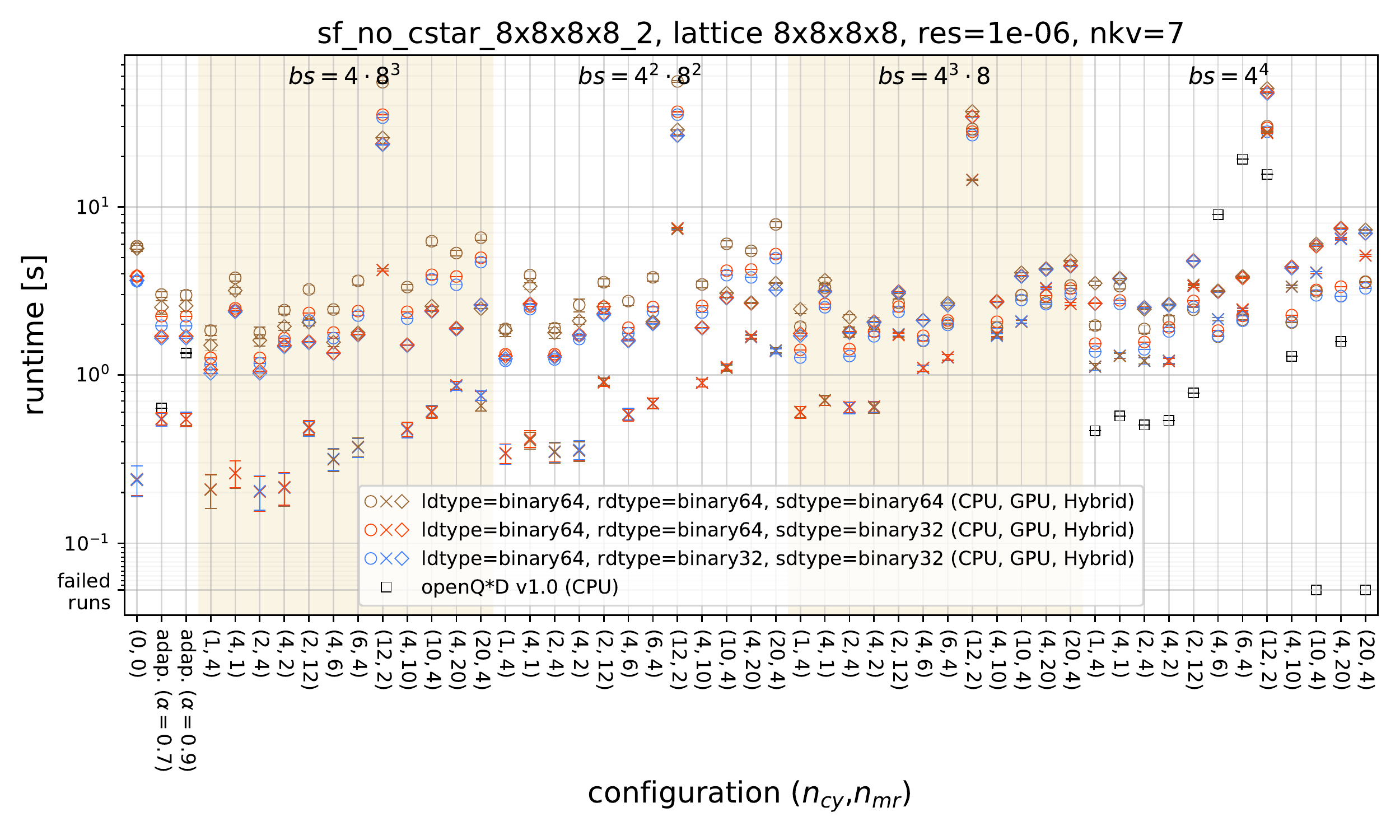}
    \caption{Time measurements for the \code{SAP\_GCR} kernel on different matrices and configurations. The measurements were conducted on an Intel(R) 6130 @ 2.10GHz with 1.5 TB memory and an NVIDIA V100 (via PCIe) GPU with 16 GB memory. The x-axis gives the configuration in terms of $(n_{cy}, n_{mr})$, the alternating shaded regions give the block size, whereas the shape of the data points indicate the processing device; circle, cross, diamond = pure CPU, pure GPU, Hybrid. Hybrid means that only the blocked problems where solved on the GPU.}
    \label{fig:sap_gcr4}
    \label{fig:sap_gcr_end}
    \label{fig:sap_gcr_sf_8x8x8x8_2}
\end{figure}

For the operator in figure \ref{fig:sap_gcr_sf_8x8x8x8_2}, we see that preconditioning gives no significant improvement. This shows that for well-conditioned operators, too much preconditioning worsens the performance. $(n_{cy}, n_{mr}) = (1,4)$ is the configuration with the least amount of preconditioning. The CPU run-time shows a strong dependence on the configuration; there are even certain \df{exceptional configurations}, eg. $(12,2)$, that are more than \num{40} times slower than the non-preconditioning case $(0,0)$. An unsuitable choice of configuration parameters can thus lead to a significant performance degradation. However, the plots show that performance of the algorithm is overly sensitive to the choice of these parameters. The adaptive variant might be of advantage here (cf. section \ref{sec:adap}).

\begin{figure}[htbp]
    \centering
    \includegraphics[width=1.0\textwidth]{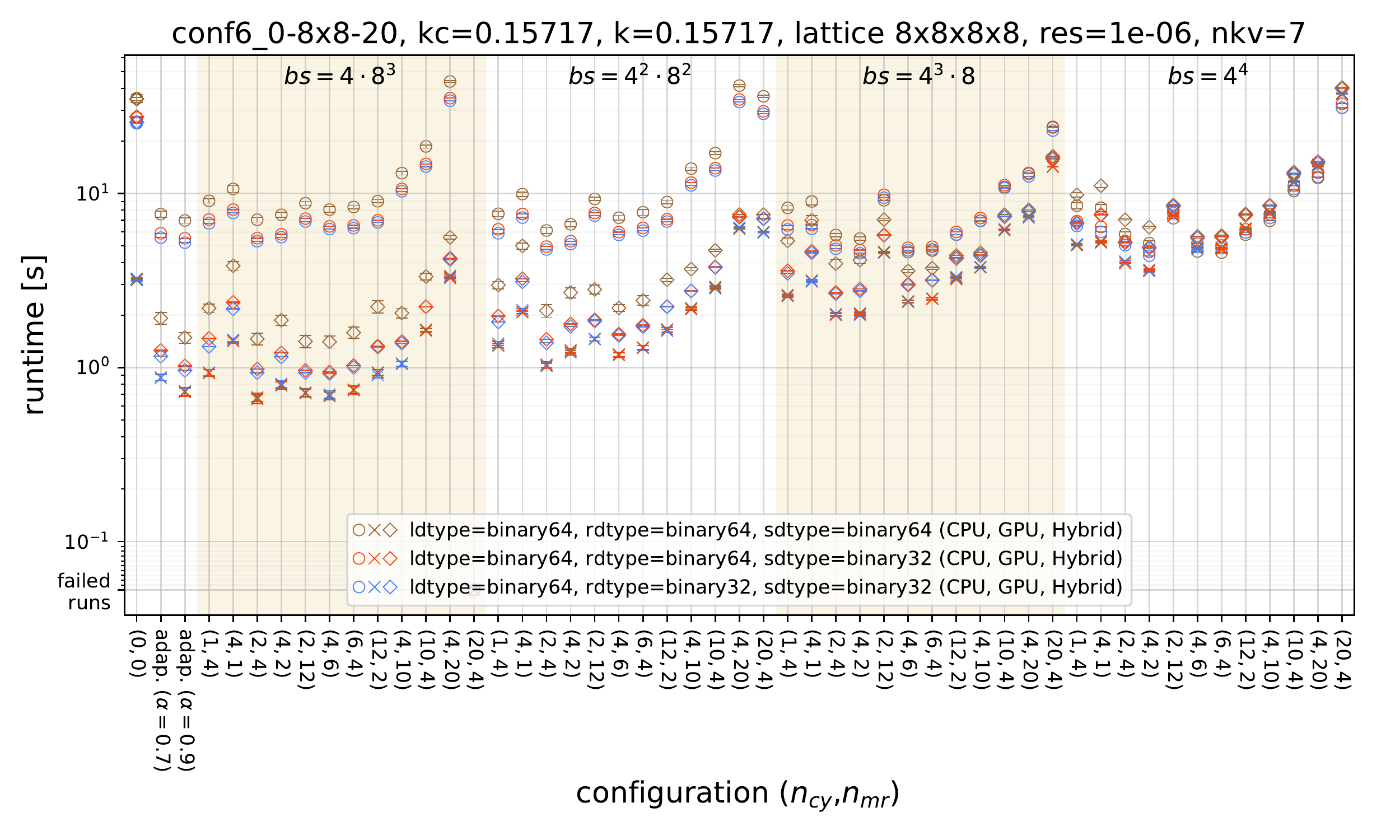}
    \caption{Time measurements for the \code{SAP\_GCR} kernel on different matrices and configurations. The measurements were conducted on an AMD EPYC 7742 CPU @ 2.25GHz with 512 GB memory and an NVIDIA A100 (via SXM4) GPU with 40 GB memory. See figure \ref{fig:sap_gcr4} for more information.}
    \label{fig:sap_gcr1}
    \label{fig:sap_gcr_conf6_0-8x8-20_0.15717}
\end{figure}

The operator in figure \ref{fig:sap_gcr_conf6_0-8x8-20_0.15717} was at the critical point $k=k_c$. This is the regime where SAP-preconditioning shows its true potential; nearly all cases performed better than the trivial case. For the pure-CPU cases, we see no strong dependence on the amount of preconditioning, but on the block size. Small block sizes seem to be beneficial, while the pure-GPU variant prefers large block sizes. The hybrid cases -- as usual in-between -- are closer to the pure-GPU ones, because despite being hybrid most of the work is done on the GPU. The pattern within a certain block size is repeating and the best amount of preconditioning seems to be at $(4,6)$.

Since the algorithm is applied to many different Dirac-operators among evolving HMC-trajectories -- some well-conditioned, some ill-conditioned -- it can be hard or even impossible to choose a set of parameters suitable for all cases. In particular, it is unavoidable to accidentally make a choice that falls on a configuration with exceptional long convergence time for a certain Dirac-operator within the long running HMC-simulation. It is therefore advisable to have the possibility to change the parameters during an active run or a configuration that adapts. This motivates the following proposal.

\subsection{Proposal for an adaptive variant of SAP+GCR}

\label{sec:adap}

Since the choice of parameters in the SAP+GCR kernel seems non-trivial, we propose an adaptive variant of this algorithm. In this version, the interpretation of the two parameters $n_{cy}$, $n_{mr}$ from table \ref{tab:sap_gcr_settings} is slightly different; they now denote the \textit{maximal} amount of Schwart-cycles and MR-steps, respectively. The actual $n_{cy}$, $n_{mr}$ were chosen automatically in every iteration anew. They were determined as follows: If -- after a Schwarz-cycle -- the norm of the residual is not lower than the residual norm before the cycle, the preconditioning phase ends. Thus, at least one Schwarz-cycle is performed in every step. A similar, but slightly more complicated strategy is applied to determine the number of MR-steps. There are \num{3} exit conditions for the MR-solver:

\begin{enumerate}[label={\arabic*)}]
  \item If -- after at least \num{4} MR-steps -- the norm of the residual on the block is larger than $\alpha = 0.9$\footnote{Ironically, the choice for the value of $\alpha \in (0, 1]$ is again non-trivial. Small values cause less preconditioning while values close to \num{1} will end up in more or even the maximal number of MR-steps. But since we want to optimise for ill-conditioned systems and the penalty for well-conditioned systems is acceptable, it is advisable to choose $\alpha$ large, such as $\alpha = 0.9$} times the previous residual norm, the MR-solver exits, and the application continues processing the next block.
  \item If the norm of the blocked residual becomes larger than the previous residual norm, the solver exits immediately, even if only one MR-step is executed.
  \item If the norm of the blocked residual is smaller than the tolerance\footnote{This is the tolerance calculated in the GCR solver divided by the number of blocks, $tol = res*\norm{\eta}/n_b$, where \code{res} is the desired relative residual given as configuration option (see table \ref{tab:sap_gcr_settings}), $\eta$ is the source vector and $n_b$ is the number of blocks.}, the algorithm exit immediately too.
\end{enumerate}

Every block is treated differently in every cycle. A maximum of \num{20} Schwarz-cycles and \num{20} MR-steps on each block would be performed if the above exit conditions never kick in. The third exit condition above makes sure to not overshoot the mark if the algorithm performs a lot of Schwarz-cycles and MR-steps, i.e. if the problem is already solved while in the preconditioning phase. This can happen if the operator is very well-conditioned or in the very last GCR-step before converging to the desired relative residual. Therefore, the adaptive version tries to find the optimal configuration for every iteration of the GCR-solver, for every Schwarz-cycle and for every block separately. By empirical observation of the results, the adaptive variant usually performs nearly maximal amounts of preconditioning in the first few GCR-steps, then rapidly decreases after some steps and finally saturate to the minimal amount that stays until convergence.

The results on how this adaptive variant competes with static configurations can be seen in the figures  indicated by a configuration $\text{adap. }(\alpha=0.9)$ and $\text{adap. }(\alpha=0.7)$. Although the adaptive variant of the algorithm is not the fastest among all configurations, the plots show that it is certainly the most versatile one. It can be of benefit if the condition of the operator is not known beforehand and might even change drastically within a long running simulation.

A reference implementation has been added to openQ$^\star$D and can be found in the GitLab repository ref. \cite{gitlab_adaptive}.

\acknowledgments

We acknowledge access to Piz Daint at the Swiss National Supercomputing Centre, Switzerland under the ETHZ's share with the project IDs s299 and c21. This work will be continued as part of the PASC project "Efficient QCD+QED Simulations with openQ$^\star$D software". Finally, we want to thank the people from NVIDIA (Mathias Wagner and Kate Clark) for interesting discussions and their useful input.

\bibliography{main}
\bibliographystyle{ieeetr} 


\end{document}